\documentclass[lettersize,journal]{IEEEtran}
\usepackage{amsmath,amsfonts}
\usepackage{array}
\usepackage[caption=false,font=normalsize,labelfont=sf,textfont=sf]{subfig}
\usepackage{textcomp}
\usepackage{stfloats}
\usepackage{url}
\usepackage{verbatim}
\usepackage{graphicx}
\usepackage{cite}
\usepackage{ragged2e}
\begin{document}

\title{Efficient onboard multi-task AI architecture based on self-supervised learning}

\author{Gabriele Inzerillo, Diego Valsesia, Enrico Magli
        % <-this % stops a space
\thanks{The authors are with Politecnico di Torino - Department of Electronics and Telecommunications,  Italy.  Email:{name.surname}@polito.it. Corresponding author: Diego Valsesia. This study was partially funded by the Italian Space Agency under ASI contract N. 2023-23-U.0 within the programme on ``Analisi dati e immagini''. We acknowledge the CINECA award under the ISCRA initiative, for the availability of high performance computing resources and support.}}% <-this % stops a space


\maketitle

\begin{abstract}
There is growing interest towards the use of AI directly onboard satellites for quick analysis and rapid response to critical events such as natural disasters. This paper presents a blueprint to the mission designer for the development of a modular and efficient deep learning payload to address multiple onboard inference tasks. In particular, we design a self-supervised lightweight backbone that provides features to efficient task-specific heads. The latter can be developed independently and with reduced data labeling requirements thanks to the frozen backbone. Experiments on three sample tasks of cloud segmentation, flood detection, and marine debris classification on a 7W embedded system show competitive results with inference quality close to high-complexity state-of-the-art models and high throughput in excess of 8 Mpx/s.  
\end{abstract}
o

\section{Introduction} \label{sec:intro}

In conventional satellite imaging systems, the satellite's task is to capture data, typically images, and transmit them to the ground segment for processing into various levels of products to be delivered to the final users. This transmission and processing chain can result in significant delays, in the order of days, to the availability of imagery to end users. This is especially undesirable in time-sensitive problems, such as natural disasters, where it is critical to obtain the data as soon as possible. 

An emerging paradigm \cite{giuffrida2021varphi} is to move part of the processing directly onboard the satellite, so that it can detect potentially critical situations in real time and trigger early warnings whose quick transmission to the ground segment is prioritized. This process requires an at least partial onboard formation of image products, followed by image analysis; a full onboard pipeline of optical and SAR image formation, analysis and alert generation has been demonstrated in \cite{kerr2021eo}. Achieving this goal requires facing a challenging tradeoff between the quality of the detection, its latency, and the computational constraints of onboard platforms dictated by the strict power budgets of satellites. Moreover, multispectral images are capable of detecting several phenomena of interest, such as floods, fires, clouds, marine debris, and many more, leading to the capability of addressing multiple tasks at the same time. Indeed, onboard multitask inference would not only allow to monitor multiple critical phenomena at the same time but also synergistically improve the entire platform. For example, segmentation of clouds could be used to optimize the onboard compression algorithm by lowering the data rate for cloud-covered areas. 

Deep learning is key to achieving state-of-the-art performance for the detection tasks we aim at solving. However, designing a mission with its onboard use with real-time performance for multiple tasks is far from trivial and requires careful study of multiple issues. In this paper, we study how a potential mission could define an AI computational payload providing low-latency responses to multiple tasks with efficient use of resources and flexible design. In particular, we envision the use of a neural network composed of a lightweight backbone to extract features from multispectral input images at multiple spatial resolutions, including relatively fine-grained ones. This feature extractor is trained in a self-supervised manner to exploit large collections of unlabeled imagery by the mission operations center, and the model is made available to entities (e.g., third party contractors), or it is made publicly available. The features are then used by lightweight neural network heads, working in parallel, each specialized for one image analysis task. These heads can be designed independently by third-party contractors, with domain knowledge of the tasks. The third-party contractors will be required to use the backbone without the ability to change its weights, so that multiple heads can share the features for their respective tasks. This approach also conveniently limits the data requirements for the third parties who have to develop the task-specific heads, since they can leverage the backbone features and only need a small amount of labels to train the small heads. Once deployed onboard, the architecture can solve as many task as the number of heads in parallel, but conveniently sharing features to significantly reduce the computational requirements. Finally, the modular approach allows in-flight updating of the backbone and heads, or even addition of new tasks.

To summarize, this paper presents the following key novel contributions:
\begin{itemize}
    \item we study how to design the onboard AI system for an Earth observation mission required to address multiple tasks, analyzing the entire framework needed to accomplish this goal, the technical details of the individual components, and presenting novel methodologies; 
    \item we show that a desirable design pretrains a backbone neural network with a self-supervised strategy, and, in contrast with classical self-supervised learning (SSL) literature, keeps it frozen to allow processing multiple tasks in parallel;
    \item we propose a novel self-supervised learning method that can extract features with both fine and coarse spatial resolution, outperforming existing SSL methods in remote sensing;
    \item we evaluate the effects of quantization on SSL pretrained models, a topic rarely explored in the SSL literature;
    \item we demonstrate a lightweight and modular design that provides inference accuracy close to that of high-complexity state-of-the-art models while achieving higher throughput on three tasks for onboard computing (clouds segmentation, floods segmentation and marine debris classification)
    \item we demonstrate that the proposed design has an excellent trade-off between quality, throughput, and power consumption on a low-power 7W embedded system.
\end{itemize}

\section{Background} \label{sec:bkg}

\subsection{Self-Supervised Learning (SSL)}

Self-Supervised Learning (SSL) has emerged in the last years as a poweful paradigm in deep learning, aiming at learning good representations that capture intrinsic data features without relying on human-labeled annotations. This is critical in remote sensing due to scarcity of labeled data and the abundance of unlabeled imagery. Contrastive learning is currently one of the most successful SSL techniques where informative representations emerge from minimizing the distance between the feature-space embedding of the same image subjected to two distinct random augmentations (positive pair) while maximizing the distance between representations derived from distinct images (negative pairs). Works such as SimCLR \cite{chen2020simple} highlight the efficacy of contrastive methods in learning resilient and generalizable representations, although they are not without flaws. In fact, they often need a large batch size to work properly and also the handling of negative pairs needs to be carefully managed. The need of a large batch size was partially solved by MoCo (Momentum Contrast) \cite{he2020momentum}, using a momentum moving average encoder. Newer methods like BYOL \cite{grill2020bootstrap} also address the problem of creating truly negative pairs by relying only on positive ones.
Furthermore, He \textit{et al.} \cite{he2018rethinking} observed that image-level learning does not always provide good representations for sub-pixel level tasks such as semantic segmentation and object detection limiting the effectiveness of SSL only to image-level tasks such as classification. For this reason, research has started investigating ``Dense SSL'' techniques \cite{islam2023selfsupervised, ziegler2022selfsupervised, wang2021dense} capable of learning more fine-grained features.

In the context of remote sensing, some works \cite{tao2020remote,kang2020deep,li2021semantic,montanaro2022semi} have explored pretext tasks and ways of framing contrastive learning that lead to SSL features that are more suitable for the remote sensing detection tasks. It is also worth noting that a typical framework for most works is to use SSL as a pretraining technique, followed by supervised finetuning of the entire model, including application head. This typically results in better accuracy than what would be obtained by keeping the backbone frozen to the SSL-trained weights. However, in this work, we will not follow this finetuning approach, as it poses undesirable restrictions in the mission design, such as the inability to develop heads independently.

\subsection{Efficient Inference with Neural Networks}

The pursuit of efficient inference in deep learning has spurred numerous innovations in the realm of lightweight networks and quantization techniques. Lightweight architectures like MobileNet \cite{howard2017mobilenets}, ShuffleNet \cite{zhang2017shufflenet}, and EfficientNet \cite{tan2020efficientnet} have aimed at reducing computational overhead while preserving accuracy. These networks employ strategies such as depth-wise separable convolutions, channel shuffling, and compound scaling to achieve a balance between model size and accuracy. On the other hand, quantization techniques, such as post-training quantization \cite{jacob2017quantization} and quantization-aware training \cite{hubara2016binarized}, aim to reduce model size and increase inference speed by representing weights and activations using lower bit precision. Additionally, methods like knowledge distillation \cite{hinton2015distilling} and neural architecture search \cite{zoph2017neural} have also been used in crafting efficient networks, either by transferring knowledge from larger models to smaller ones or by automating the design process to discover architectures optimized for fast inference.

\subsection{Onboard AI-Based Processing}

In recent years, the advancement of neural networks and AI has extended to onboard satellite processing systems and edge-devices in general, enabling real-time data analysis directly in space. This advancement is particularly significant in the context of remote sensing, where the ability to process large volumes of onboard imagery can dramatically reduce latency, optimize bandwidth utilization, and enable more responsive and autonomous satellite operations. However, onboard AI is faced with several challenges in the design of lightweight and power-efficient systems.
Several studies\cite{ai_on_board} have started demonstrated the feasibility and effectiveness of AI-based onboard processing for remote sensing tasks. Yao \textit{et al.} \cite{yao2019board} was one of the first works to address the challenge of running deep learning models directly onboard satellites, proposing a simple framework for ship detection on small satellites. Notably, Giuffrida \textit{et al.} \cite{on-board_demonstrator} demonstrated the integration of AI for onboard data processing in real Earth Observation missions ($\Phi$-Sat-1 by the European Space Agency), showcasing the feasibility of running deep convolutional neural networks on the Intel Movidius Myriad 2 hardware accelerator for real-time cloud detection on hyperspectral images. Ziaja \textit{et al.} \cite{ziaja2021benchmarking} proposed and extensive benchmark of various deep learning models on edge devices for onboard space applications. Růžička \textit{et al.} \cite{ruuvzivcka2022ravaen} introduced a lightweight model for change detection onboard satellites based on Variational Auto-Encoders.

In this paper, as a demonstration of the performance of the proposed design in a low-power setting, we test using an Nvidia Jetson Orin Nano, a Commercial-Off-The-Shelf (COTS) hardware platform. This should be considered as a low-power demonstrator, and not necessarily representative of a real flight implementation. Indeed, we recognize that different space missions may choose different approaches to the integration of deep learning in the onboard computing platforms depending on specific mission characteristics. For instance, they may rely solely on FPGAs, or a combination of FPGAs with GPUs/CPUs or even just a COTS System-on-Chip.

\section{Method} \label{sec:method}

In this section, we introduce a novel, modular and lightweight multitask architecture tailored for usage onboard satellites for low-latency inference. We also go beyond the mere architecture design by presenting ideas that serve as a blueprint for a mission planning, which in turn affect decisions about the neural network development.

\subsection{Mission Vision}
The approach towards the design of neural network components described in later sections stems from ideas about the specific goals and planning requirements of a hypothetical mission. 
We envision a Sentinel2-like multispectral imager with additional capabilities provided by onboard AI.
In particular, the first novel capability would be AI-assisted onboard compression. It is known that image compression methods, including existing standards for hyper- and multispectral images, can be aided by cloud detection \cite{cilia2020onboard} to provide pixel-level maps of regions where compression quality can be lowered to significantly save data rate. Cloud segmentation is therefore a desirable task to be included for any onboard AI capability. Furthermore, the second capability of interest is the generation of alerts to be delivered to the ground segment with low latency when specific phenomena, such as natural disasters, are detected. For this capability, it is desirable to produce both pixel-level segmentation maps (e.g., to detect the extent of flooded areas) as well as whole-image classification labels (e.g., to detect debris presence and its type, or presence of active fires). These requirements clearly outline the need to have features with fine spatial granularity so that the segmentation tasks can be solved effectively.  

\begin{figure}[t]
\centering
\includegraphics[width=0.99\columnwidth]{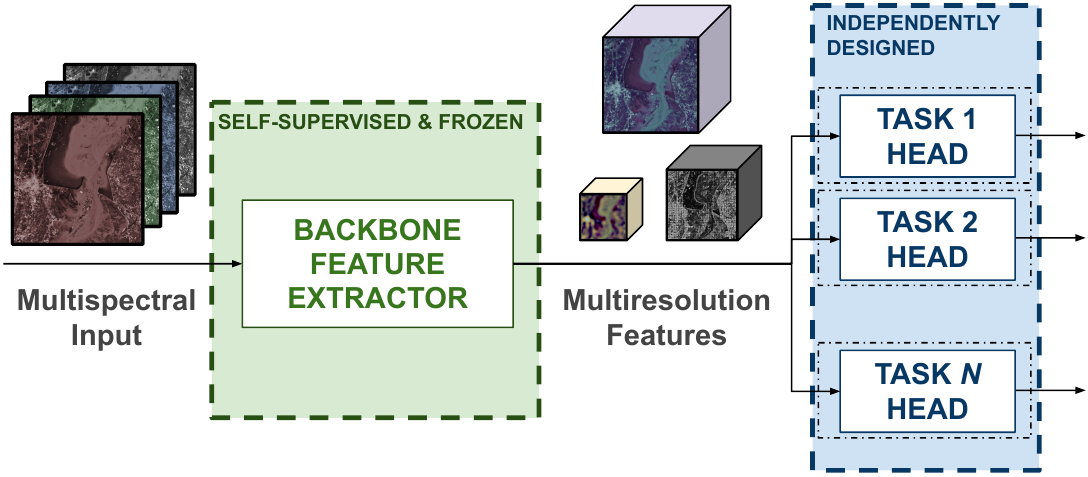}
\caption{High-level design for modular multitask neural network design. A lightweight backbone is trained with SSL and then frozen to generate universal standard multiresolution features. Application-specific heads can be independently developed to exploit such features for inference tasks.} 
\label{fig:3.1}
\end{figure}

Concerning mission planning, a modular approach is required so that multiple parties can cooperate in the design of the AI module and its possible update. In particular, Fig. \ref{fig:3.1} highlights multiple modules to be developed independently. A backbone serves as a universal feature extractor. This is developed independently of the specific tasks to be solved, except for the requirement of providing features with fine-grained spatial resolution. The features extracted by the backbone for a specific input are then used by task-specific heads which are comparatively smaller neural networks. These can be assigned to multiple third-party domain experts for their development. However, in order to guarantee reusability of the features for all tasks, such third parties are not allowed to fine-tune the backbone.

Finally, the entire neural model must be lightweight, so that it can fit the limited memory of embedded systems and provide a high enough throughput. Targets for throughput depend on the specific mission requirements in terms of coverage and latency. However, as a rough idea, we can consider as generally adequate a throughput in the same order of magnitude of that of the image compression subsystem which is designed to keep up with the satellite acquisition. This is typically in the tens of millions of samples per second \cite{santos2015multispectral,fjeldtvedt2018efficient} (with a spectral vector composed of one sample for each band), so we can consider as more than satisfactory a neural network labeling one to ten million spatial locations per second.

\subsection{Self-supervised Backbone}\label{subsec:backbone}

\begin{figure}[t]
\centering
\includegraphics[width=0.99\columnwidth]{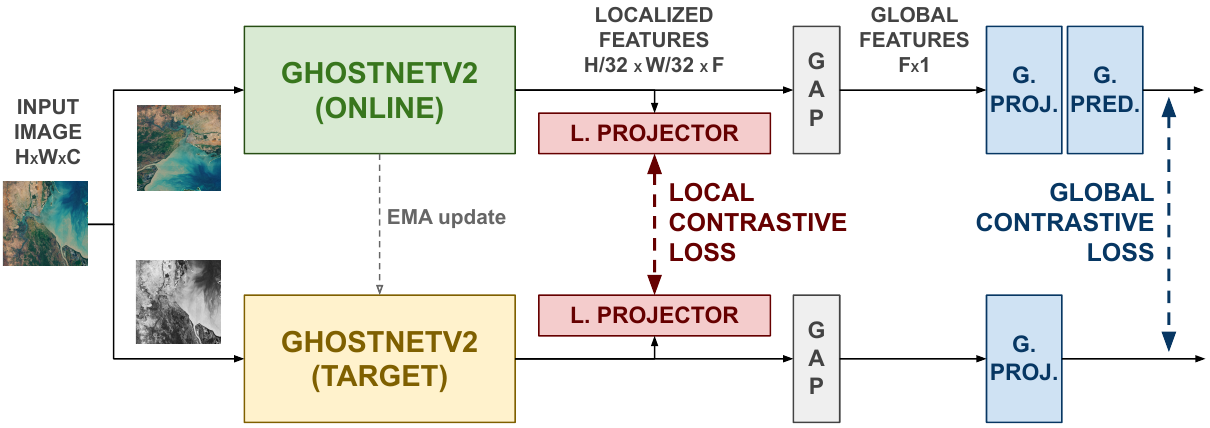}
\caption{SSL training of the GhostNetV2 backbone using Local Contrastive Loss. In this framework, the online branch (upper) receives an input view of the image generated solely by applying spatial transformations, while the target branch (lower) processes a second view created by applying both spatial and intensity transformations. The structural configuration of the framework mirrors that presented in BYOL \cite{grill2020bootstrap}, augmented by the inclusion of a local contrastive branch. GAP is global average pooling.}
\label{fig:3.2}
\end{figure}

The main component of the neural network architecture is the backbone, which acts as a universal feature extractor. This feature extractor comprises a deep neural network that takes as input a multispectral image and computes a semantically meaningful representation composed of a number of features. This representation is then leveraged by further task-specific neural network heads for various applications. 

The backbone feature extractor is designed and trained to produce features that can effectively be shared by all the heads; this means that the representations produced by the backbone must be general enough to adapt to a variety of possible vision tasks (such as classification, semantic segmentation, object detection and more). This approach ensures two fundamental aspects within our architecture:

\begin{itemize}
  \item \textit{Modularity}: the backbone is task-agnostic, thus it operates independently of the specific application heads or tasks we incorporate. This independence allows for separate training and functioning, promoting a modular framework where components can be adjusted or added without extensive restructuring.
  \item \textit{Efficiency}: the computational complexity is primarily concentrated within the backbone, performing the most resource-demanding computations just once. Subsequently, each head can execute its task in parallel, using these pre-processed features. This parallel execution enhances overall efficiency by minimizing redundant computations and optimizing task-specific processing.
\end{itemize} 

In principle, an ideal backbone would be constituted by a foundational model \cite{awais2023foundational} trained on vast amounts of data to generate highly general representations. While such foundational models are starting to emerge in the remote sensing literature \cite{jakubik2023foundation}, having one that is also lightweight remains elusive. One path towards a model of this kind is the use of SSL techniques which can exploit large datasets of unlabeled imagery and produce task-agnostic representations, coupled with an efficient design to match current computational capabilities of embedded systems.

% SSL
Concerning SSL training, in this paper we chose to use the 590,326 Sentinel-2 images from the BigEarthNet dataset\cite{Sumbul_2019}. We propose to use a SSL technique that adapts the methodology outlined in BYOL \cite{grill2020bootstrap}, with a \textit{Local Contrastive Loss} inspired by the work of Islam \textit{et al.} \cite{islam2023selfsupervised} to promote spatial features with a fine-grained resolution, useful for pixel-level tasks such as segmentation. A high-level overview of SSL training is depicted in Fig. \ref{fig:3.2}. Two augmentations of an input multispectral image go through the online (top) and the target (bottom) networks, the latter being composed of weights obtained from a moving average of the weights of the online network. The projector layer is a linear operation on a spatially-pooled representation of the entire image in a feature space. The online network has an extra linear layer called predictor. A global contrastive loss minimizes a dissimilarity metric between the output of the online predictor and the target projector. This global loss ensures that representation are globally semantically informative, and promotes clustering according to semantic classes for whole-image classification problems. However, it is not sufficient to ensure that the backbone learns fine-grained spatial features for segmentation problems. This is why a local contrastive loss is used to minimize pixelwise feature dissimilarity before spatial pooling. Overall, the SSL training loss is thus:
\begin{align}
    L = \lambda L_{LC} + (1 - \lambda)  L_{GC}
\end{align}
with tradeoff parameter $\lambda$ and
\begin{align}
    L_{LC} &= \frac{1}{|P|} \sum_{(p_c,p_{sc}) \in P} NLL(p_c,p_{sc}) \label{eq:LLC} \\
    L_{GC} &= 2-2\cdot\frac{\langle q_\theta(z_\theta), z'_\xi \rangle}{||q_\theta(z_\theta)||_2\cdot||z'_\xi||_2 }
\end{align}
with $q_\theta(z_\theta)$ the predictor's output of the online (upper) network and $z_\theta$ the output of the projection of the target network in Fig. \ref{fig:3.2}. In Eq. \eqref{eq:LLC} $p_c$ are known points selected by defining a uniform $h\times w$ 2D grid in the image that was augmented only by applying color transformations. Having defined the grid, and thus the points, and knowing the spatial transformations applied to the other image we can obtain the corresponding $p_{sc}$ points in the second image, thus creating a $p_c-p_{sc}$ point mapping between the pair of images. We use the latter to compute the Negative Log-Likelihood as follows:
\begin{align}
    NLL(p_c,p_t)=-\log\frac{e^{C'(p_c,p_{sc})/\tau}}{\sum_{k\in\Omega_{sc}}e^{C'(p_c,p_{k})/\tau}}
\end{align}
where $\Omega_{sc}$ is the set of points $p_{sc}$, $\tau$ is a temperature hyperparameter and $C' \in R^{(h\times w)\times(h \times w)}$ is a dense correspondence map between the mapped points in the two images:

\begin{align}
    C'(p_c,p_{sc}) &= \frac{F_\xi(p_c)^T F_\theta(p_{sc})}{||F_\xi(p_c)||_2 ||F_\theta(p_{sc})||_2 } \label{eq:CorrespondanceMap} 
\end{align}
where $F_\xi(p_c)$ and $F_\theta(p_{sc})$ are the dense feature representation of the points obtained, respectively, from the target and online networks.

% architecture
Concerning the architecture design, we suppose that four spectral bands (Red, Green, Blue and Near Infrared) are used as input. The choice of bands is a tradeoff between the tasks to be solved and computational complexity: using all the available spectral bands could provide a more flexible backbone, but also increase computational complexity. For the sample tasks explored in this paper, the RGB and NIR bands provide adequate information, but other tasks could require additional bands. For instance, SWIR could be useful to target fire detection and one could imagine to implement a slightly more complex head that takes as input the backbone features as well as the pixels of a new (e.g. SWIR) channel, and combines them to produce output for this specific task.  

Given our primary goal of designing an efficient network, our approach to creating the backbone feature extractor is based on GhostNetV2 \cite{tang2022ghostnetv2}.
GhostNetV2 is a state-of-the-art lightweight convolutional neural network (CNN), specifically designed for fast inference on mobile and edge devices. As reported in \cite{tang2022ghostnetv2}, its performance surpasses that of MobileNetV3-L \cite{mobilenetv3} by approximately 1\%, achieving a top-1 accuracy of 77.8\% in ImageNet classification. Notably, this achievement comes with a slightly increased number of FLOPs compared to MobileNetV3's 355 MFLOPs. Moreover, GhostNetV2 exhibits superior performance by approximately 2\% over MobileViT-XS \cite{mobilevit}, despite MobileViT-XS has almost twice as many FLOPs as GhostNet.
The main innovation of GhostNetV2 lies in realizing that conventional CNNs have highly redundant feature maps. Therefore, they can be obtained in a less expensive manner by initially generating a set of intrinsic feature maps, and then using multiple cheap linear operations on them to derive the remaining redundant feature maps. This goal is achieved by a structure called GhostNet Bottleneck, comprising stacked GhostNet modules, each incorporating a ``hardware-friendly'' attention mechanism known as Decoupled Fully Connected Attention. This attention mechanism aims to create feature maps that incorporate both local and long-distance information. Due to its extreme efficiency, combined with excellent performance, we selected GhostNetV2 as the backbone for feature extraction, excluding the four final layers specifically designed for classification. The output features to be used for the task-specific heads are taken at multiple depths of the GhostNetV2 architecture in order to provide a multiresolution feature bank which is known \cite{sun2019highresolution,valsesia2022super} to be more effective than a single resolution for certain tasks. Specifically, feature maps after the 5th, 7th, and 10th layers at $\frac{1}{8}$, $\frac{1}{16}$ and $\frac{1}{32}$ of the input spatial resolution are selected.

Lastly, we want to emphasize that the choice of GhostNetV2 is motivated by being the state-of-the-art model among low-complexity backbones at the time of writing, presenting an excellent tradeoff between complexity and accuracy, which allows us to verify if the onboard multitask AI system can achieve good performance. However, the considerations in this paper are more general and GhostNetV2 could be replaced with any backbone providing multiresolution features, resulting in different tradeoffs between accuracy, latency and power consumption.

\subsection{Task-specific Heads}\label{subsec:heads}

The general framework outlined in this paper enables a large variety of applications to be addressed thanks to the features extracted from the backbone. In order to evaluate the effectiveness of our design, we tested three tasks which can be relevant for onboard inference: cloud cover segmentation \cite{cloudcover}, floods segmentation \cite{s1floods11}, and marine littering whole-image multi-label classification \cite{marida}. 
The following datasets have been used for the sample tasks.

\begin{itemize}
    \item \textit{Sentinel-2 Cloud Cover Segmentation Dataset} \cite{cloudcover}: the dataset, developed by the Radiant Earth Foundation\footnote{https://radiant.earth/}, comprises 22,728 Sentinel-2 satellite images and their corresponding binary cloud masks. Each image has $512\times512$ pixels and represents imagery of a distinct area captured at a specific instance.
    \item \textit{Sen1Floods11} \cite{s1floods11}: this dataset encompasses images from both Sentinel-1 and Sentinel-2 satellites, featuring binary masks distinguishing permanent water bodies from water associated with flood events. Focusing specifically on multispectral imagery, we filtered the dataset to retain solely the multispectral L1C images from Sentinel-2. This subset comprises only 446 images, each having $512\times512$ pixels.
    \item \textit{MARIDA} \cite{marida}: Marine Debris Archive (MARIDA) is a dataset for the classification of marine debris. The dataset includes 1381 Sentinel-2 multispectral images of 256x256 pixels, which distinguishes marine debris from various coexisting marine classes, including Sargassum macroalgae, ships, natural organic material, waves, wakes, foam, different water types (e.g., clear water, turbid water, sediment-laden water, shallow water), and clouds. We use this dataset for whole-image multi-label classification where the entire input image is classified in one of 11 classes (different water types are aggregated into one class, as in the original paper, reducing the number from the original 15 to 11). 
\end{itemize}

Three heads are therefore used in parallel in this example of multitask onboard inference. Since different kinds of tasks are to be solved, we designed two distinct low-complexity head types: one for multi-label classification and the other for segmentation. 

\subsubsection{Classification Head}
the classification head consists of the four layers removed from the GhostNetV2 backbone, as detailed in Subsection \ref{subsec:backbone}. It serves as a straightforward neural network comprising an initial global average pooling layer of only the features at the coarser spatial resolution, followed by a fully-connected layer with Rectified Linear Unit (ReLU) activation and a fully-connected layer with Softmax activation.

\subsubsection{Segmentation Head} 
\begin{figure}
\includegraphics[width=0.99\columnwidth]{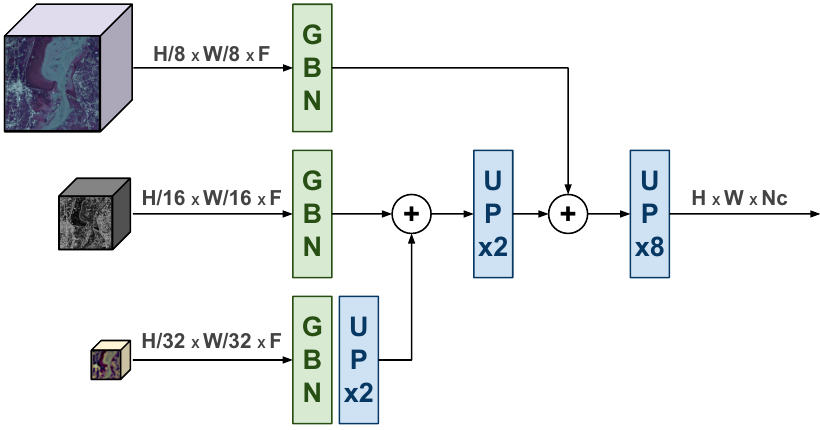}
\caption{Architecture of a Segmentation Head exploiting multiresolution features extracted by the backbone. UP is bilinear upsampling, GBN is the GhostNetV2 Bottleneck Module \cite{tang2022ghostnetv2}.}
\label{fig:3.3}
\end{figure}

our segmentation head shares similarities with FCN-8s \cite{shelhamer2016fully} or HRNet \cite{SunXLW19, WangSCJDZLMTWLX19, sun2019highresolution}, as we adopted a similar multiresolution approach of aggregating feature sets extracted from various layers of the backbone. An overview is shown in Fig. \ref{fig:3.3}. Following the efficiency paradigm, we utilized the GhostNet Bottleneck module instead of traditional 2D convolutions to reduce the number of parameters and FLOPs. This module is applied in parallel to the three resolutions which are then added as residuals after bilinear upsampling.

The modularity of our architecture is evident: the backbone generates flexible, generalized features suited for multiple heads, each requiring fine-tuning solely on task-specific datasets. Introducing a novel task distinct from prior ones seamlessly integrates through the creation of a corresponding new head and its subsequent fine-tuning. Notably, fine-tuning the entire architecture (i.e., backbone and a single application head) is not desirable since the selective fine-tuning of the application heads, while keeping the backbone frozen, preserves the architecture's modularity and allows multiple tasks to be performed in parallel by the different heads. Furthermore, this approach requires a minimal amount of data for training, ensuring extremely fast fine-tuning of application heads.

\subsection{Quantization}

In addition to designing lightweight modules, neural network quantization \cite{jacob2017quantization} is also critical to improve memory requirements and inference speed. Our architecture underwent a post-training quantization process employing an 8-bit integer precision (INT8) scheme. In particular, static quantization is used to quantize both weights and activations and perform fully-integer inference. INT8 is particularly optimized for implementation on embedded devices and, generally, but not always, suffers from small penalties in inference accuracy.
A different, task-specific calibration dataset has been used for quantization of each head.

\section{Experimental results}  \label{sec:experiments}

In this section, we initially validate the proposed design against state-of-the-art methods comparing their complexity as well as inference performance (in both FP32 and INT8 precision) on the three sample tasks presented in Sec. \ref{subsec:heads}. Then, we use low-power hardware to analyze how the propose design fares in terms of total energy consumption and throughput. Lastly, we present ablations by comparing a set of experiments featuring backbones of varying sizes and comparing how different SSL techniques affect the performance.

\subsection{Implementation details}

The SSL training of the backbone spanned 550 epochs, employing a learning rate of $3 \times 10^{-4}$ and a batch size of 3400 parallelized over 4 Nvidia A100 GPUs. After the completion of SSL training, we extracted the GhostNetV2 from the online branch and employed it as backbone feature extractor.
Following a procedure similar to that presented in \cite{islam2023selfsupervised}, we generated two distinct augmented views of a single image: $I_s$ resulted solely from spatial transformations encompassing horizontal and vertical flips, random rotations (90°, -90°, 180°, -180°), and random cropping. While, $I_c$ was derived by applying a combination of spatial and spectral transformations, including color jittering, intensity manipulation, gaussian blurring, and solarization. 
We determined that hyperparameter $\lambda=0.1$ offered the optimal balance between the global and local contrastive losses. This choice resulted in a balanced performance across sub-pixel level and image-level tasks. Unless otherwise stated, the GhostNetV2 architecture uses multiplier $\alpha=1.6$ in the choice of number of features. Finally, we specify that in order to perform network quantization in a fair way, the Intel Neural Compressor \cite{IntelNeuralCompresso} library was used for all models. Each model was quantized in the same way, using a \textit{static} configuration for post-training quantization, with arithmetic entirely on 8-bit integers and calibrating each model with a calibration sample of 500 elements.

\subsection{Model Comparison and Performance Evaluation}\label{subsec:performances}

\begin{table}
    \centering
    \setlength{\tabcolsep}{4pt}
    \caption{Model complexity ($512\times512\times4$ input)}
    \label{tab:4.1}
    \begin{tabular}{c c c c}
    \textbf{Architecture} & \textbf{Params} & \textbf{MACs} & \textbf{FLOPs} \\
    \hline
    \hline\\[-5pt]
    \multicolumn{4}{c}{\textbf{Single-task Segmentation}}\\[3pt]
    \hline
    \textbf{GhostNetv2 + Segmentation Head} & 9.55M & 2.18G & 4.47G \\
    \hline
    DeepLabV3 \cite{chen2017rethinking} + MobileNetV3-L \cite{mobilenetv3} & 11.02M & 9.83G & 19.74G \\
    \hline
    HRNet18 \cite{SunXLW19, WangSCJDZLMTWLX19, sun2019highresolution} & 9.64M & 18.39G & 37.01G \\
    \hline
    UNet \cite{ronneberger2015unet} + MobileViT-S \cite{mobilevit} & 8.04M & 18.7G & 37.63G \\
    \hline\\[-5pt]
    \multicolumn{4}{c}{\textbf{Single-task Classification}}\\[3pt] % RICALCOLARE PER 512
    \hline
    \textbf{GhostNetv2 + Classification Head} & 9.89M & 2.09G & 4.29G \\
    \hline
    MobileNetV3-L \cite{mobilenetv3} & 2.97M & 1.12G & 2.31G \\
    \hline
    ResNet50 \cite{he2015deep} & 23.52M & 21.56G & 43.29G \\
    \hline
    ViT B-16 \cite{dosovitskiy2021image} & 86.61M & 107.23G & 214.75G \\
    \hline\\[-5pt]
    \multicolumn{4}{c}{\textbf{Multitask}}\\[3pt]
    \hline
    \textbf{Proposed multitask} & 10.69M & 2.28G & 4.66G \\
    \hline
    \end{tabular}
\end{table}

\begin{center}
\begin{table*}
    \centering
    \caption{Cloud Cover Segmentation Performance Comparison}
    \label{tab:4.3}
    \setlength{\tabcolsep}{2pt}
    \begin{tabular}{c c c c c}
    \hline
    \textbf{Architecture} & \textbf{mIoU (FP32)} & \textbf{mIoU (INT8)} &\textbf{mF1 (FP32)}  & \textbf{mF1 (INT8)}  \\
    \hline
    \hline
    \textbf{GhostNetv2 + Segmentation Head (SSL) }& 82.33 & 81.7 & 88.16 & 87.67\\
    \hline
    GhostNetv2 + Segmentation Head (SL) & 83.75 & 83.41 & 89.13 & 88.85\\
    \hline
    DeepLabV3 + MobileNetV3-L & 83.47 & 81.8 & 88.95 & 87.72 \\
    \hline
    HRNet18 & 84.67 & 84.57 & 89.86 & 89.79 \\
    \hline
    UNet + MobileViT-S & 83.55 & 50.74 & 89.04 & 58.02 \\
    \hline
    \end{tabular}
\end{table*}
\end{center}

The tasks outlined in Sec. \ref{subsec:heads} include two semantic segmentation tasks (cloud cover segmentation \cite{cloudcover}, flood segmentation \cite{s1floods11}) and one classification task (marine litter \cite{marida} multi-label classification). Accordingly, we select some baselines and state-of-the-art models for segmentation and classification as terms of comparison. In particular, concerning segmentation we select DeepLabV3 \cite{chen2017rethinking} with MobileNetV3-L \cite{mobilenetv3} backbone for a well-known and efficient baseline, HRNet18 \cite{sun2019highresolution} as the high-complexity state-of-the-art model, and UNet \cite{ronneberger2015unet} with MobileVit-S as backbone \cite{mobilenetv3} as a recent approach leveraging the representational power of Transformers, albeit with an eye to complexity. Concerning classification, we consider ResNet50 \cite{he2015deep} as a standard baseline, MobileNetV3 \cite{mobilenetv3} as a lightweight method and ViT B-16 \cite{dosovitskiy2021image} as a high-complexity state-of-the-art model.

A summary of the computational complexity of various methods is presented in Table \ref{tab:4.1}. For ease of comparison, we also include the proposed design with a single head alongside the multitask case. The proposed design demonstrates a very low number of FLOPs while maintaining a comparable number of parameters relative to other models, except for the large classification models, which generally require a larger number of parameters. In the context of classification, one might argue that the proposed method has more FLOPs than MobileNetV3, however, when considering the multitask scenario, including also inference for segmentation tasks, the efficient DeepLabV3 + MobileNetV3 framework has more FLOPs than our architecture. Thus, in the multitask setting, our proposed method remains the most efficient. Regarding the complexity of the individual application heads (excluding the backbone): our single segmentation head has 401K parameters and requires 92K MACs, with 187M FLOPs for inference on a $512\times512\times4$ image provided to the backbone. Meanwhile, a single classification head has 743K parameters and requires 742K MACs, with 1.88M FLOPs for inference on a $512\times512\times4$ image provided to the backbone.

\begin{center}
\begin{table*}
    \caption{Floods Segmentation Performance Comparison}
    \label{tab:4.4}
    \setlength{\tabcolsep}{3pt}
    \begin{center}
    \begin{tabular}{c c c c c}
    \hline
    \textbf{Architecture} & \textbf{mIoU (FP32)} & \textbf{mIoU (INT8)} &\textbf{mF1 (FP32)}  & \textbf{mF1 (INT8)} \\
    \hline
    \hline
    \textbf{GhostNetv2 + Segmentation Head (SSL)} & 40.32 & 39.78 & 50.45 & 49.66\\
    \hline
    GhostNetv2 + Segmentation Head (SL) & 42.95 & 42.76 & 53.71 & 53.26\\
    \hline
    DeepLabV3 + MobileNetV3-L & 41.03 & 34.32 & 51.39 & 43.62 \\
    \hline
    HRNet18 & 54.68 & 54.62 & 65.72 & 65.51 \\
    \hline
    UNet + MobileViT-S & 59.65 & 14.75 & 70.25 & 19.18\\
    \hline
    \end{tabular}
    \end{center}
\end{table*}
\end{center}

Heads were trained on a supervised way on each task dataset, without finetuning the backbone. In order to provide a fair comparison, the other methods were pretrained on ImageNet \cite{ILSVRC15} and finetuned on the task datasets. We remark that freezing the backbone as dictated by our design goals is nevertheless penalizing compared to full finetuning. For this reason, we also report a benchmark in which the proposed model is fully finetuned for a specific task (this will be marked in the following as ``SL'' - supervised learning, in contrast to the ``SSL'' configuration for the frozen backbone), after the SSL pretraining.

Note that all tests conducted in the following section were performed using the original splits provided by the datasets to ensure that the results are comparable with those reported in the datasets' papers and associated benchmarks. The metrics presented in the tables below were calculated on the test set when available; otherwise, they were calculated on the validation set if the test set was not available in the dataset. The dataset configurations are as follows:

\begin{itemize}
    \item \textit{Sentinel-2 Cloud Cover Segmentation Dataset}: consists of 22,728 total images, with 11,748 in the training set and the remaining 10,980 in the test set.
    \item \textit{Sen1Floods11}: consists of 426 total images, with 256 in the training set, 86 in the validation set, and 89 in the test set.
    \item \textit{MARIDA}: consists of 1,381 images, with 694 in the training set, 328 in the validation set, and 359 in the test set.
\end{itemize}

\begin{figure*}[t]
\centering
% First row of images
\includegraphics[width=0.13\linewidth]{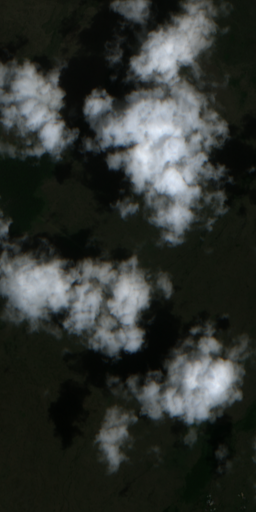}
\includegraphics[width=0.13\linewidth]{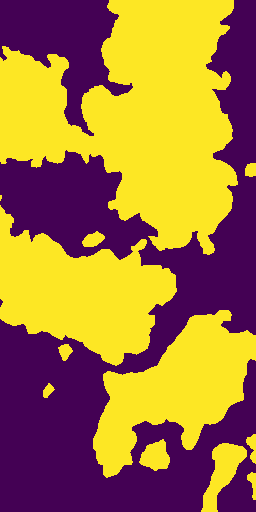}
\includegraphics[width=0.13\linewidth]{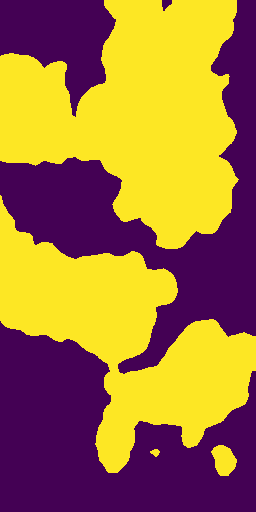}
\includegraphics[width=0.13\linewidth]{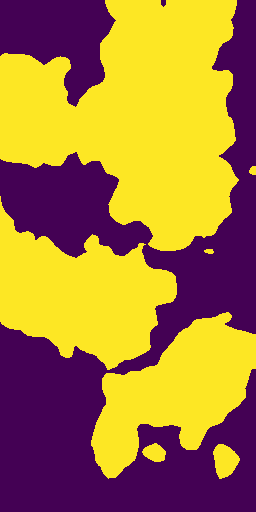}
\includegraphics[width=0.13\linewidth]{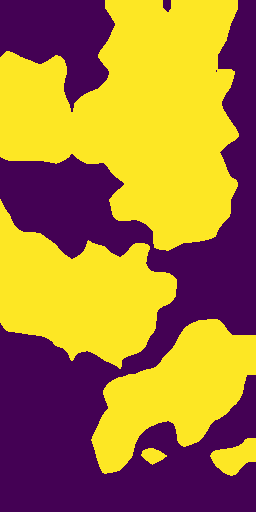}
\includegraphics[width=0.13\linewidth]{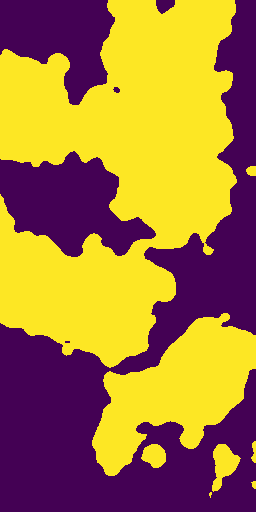}
\includegraphics[width=0.13\linewidth]{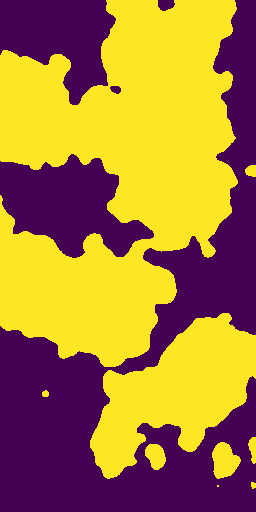} \\[4pt]

% Second row of images
\includegraphics[width=0.13\linewidth]{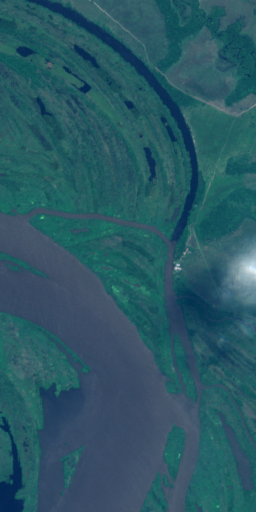}
\includegraphics[width=0.13\linewidth]{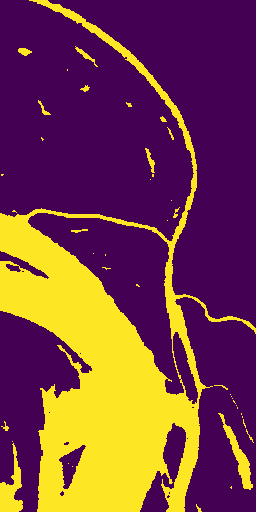}
\includegraphics[width=0.13\linewidth]{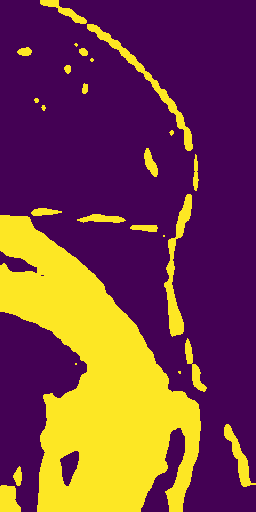}
\includegraphics[width=0.13\linewidth]{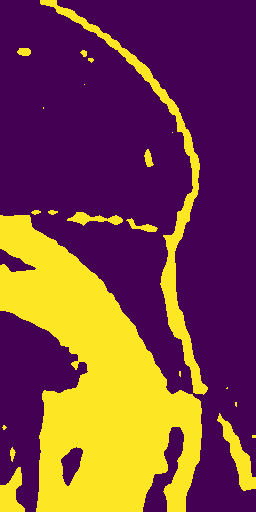}
\includegraphics[width=0.13\linewidth]{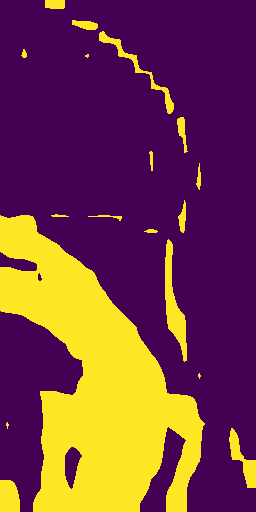}
\includegraphics[width=0.13\linewidth]{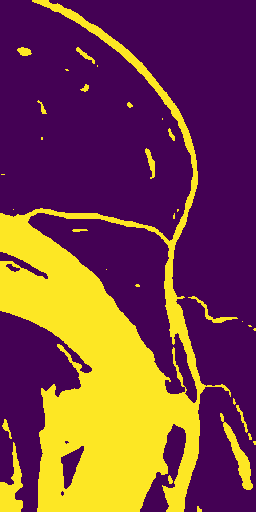}
\includegraphics[width=0.13\linewidth]{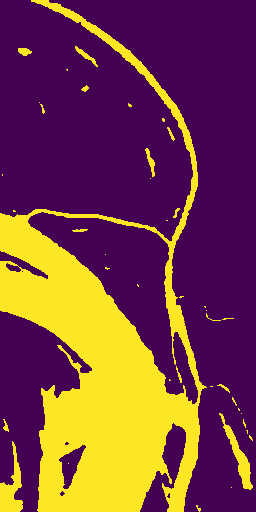}

\caption{Qualitative comparison of different segmentation models. Top row: cloud cover segmentation; Bottom row: flood segmentation. Left to right: RGB image, Ground Truth, GhostNetv2 + Segmentation Head (SSL), GhostNetv2 + Segmentation Head (SL), DeepLabV3 + MobileNetV3-L, HRNet18, UNet + MobileViT-S. }
\label{fig:seg_maps}
\end{figure*}

Tables \ref{tab:4.3}, \ref{tab:4.4}, and \ref{tab:4.5} present performance comparisons for cloud cover segmentation, flood segmentation, and marine litter classification, respectively, while in Fig. \ref{fig:seg_maps} qualitative results are shown for cloud and flood segmentation tasks, comparing the different segmentation maps obtained from the models with ground truth and the corresponding RGB image. We evaluate the two segmentation tasks using the Binary Intersection-over-Union and the F1-Score, while for the classification task, we compute only the F1-Score. We chose the Binary IoU (hereafter referred to as ``mIoU'' for brevity) to ensure consistency with published results for cloud segmentation \cite{cloudcoverchallenge} and flood segmentation \cite{s1floods11}, aligning with the official metric used in both datasets. Additionally, we compute the F1-Score for segmentation to provide a more detailed assessment of the model's performance across individual classes. In a sensitive task such as flood segmentation, it is particularly important to properly weigh the presence of false-positive pixels incorrectly labeled as ``Flood''. A significant number of false positives could suggest a nonexistent flood zone, potentially leading to unwanted triggers. Thus, relying solely on mIoU does not provide sufficient information about false positives, making the F1-Score a important complementary metric.

As shown in Table \ref{tab:4.3}, it is noteworthy that all models exhibit relatively similar performance in clouds segmentation, within about 2 percentage points of variation in mIOU between the best and worst, both in FP32 and INT8 quantization, with the exception of UNet + MobileViT-S which suffers greatly from quantization. Furthermore, it is interesting to notice that the SL version of our architecture is only marginally better than the SSL version, suggesting that the design constraint of freezing the backbone may not have a big impact. The differences in performance between the models is even smaller if we look at the F1-Score, indicating again how weel-known SOTA models with higher performance do not have excessive gains in quality metrics compared to more efficient models such as MobileNetV3 and our proposed architecture.

The results on floods segmentation, shown in Table \ref{tab:4.4}, show that the higher complexity models are generally superior in this specific task, while the proposed architecture provides better performance than the direct low-complexity alternative (DeepLabV3 + MobileNetV3-L). Indeed, scaling experiments reported in Table \ref{tab:modelsize} suggest that a larger model would improve performance in exchange of speed.

Lastly, the marine littering task addressed a problem of multi-label classification instead of segmentation, providing an analysis of how well the proposed model can address heterogeneous tasks that both need fine and coarse grained features. In this task, we notice that the proposed method is very close to the best FP32 results, and it is the best overall in INT8. It is worth remarking that the MARIDA dataset is very small, leading some highly-complex methods to overfit when finetuned.

\begin{table}
    \centering
    \caption{Marine Littering Classification Performance Comparison.}
    \label{tab:4.5}
    \setlength{\tabcolsep}{3pt}
    \begin{tabular}{c c c}
    \hline
    \textbf{Architecture} & \textbf{mF1 (FP32)} & \textbf{mF1 (INT8)} \\
    \hline
    \hline
    \textbf{GhostNetv2 + Classification Head (SSL)} & 68.03 & 63.98\\
    \hline
    GhostNetv2 + Classification Head (SL) & 61.24 & 60.38\\
    \hline
    MobileNetV3 Large & 71.94 & 50.97 \\
    \hline
    ResNet50 & 70.75 & 33.88 \\
    \hline
    ViT B-16 & 64.22 & 63.74 \\
    \hline
    \end{tabular}
\end{table}

\begin{center}
\begin{table*}
    \centering
    \caption{Throughput and Power Consumption on low-power hardware.}
    \label{tab:speed_jetson}
    \begin{center}
    \begin{tabular}{c c c c c c c}
    \hline
    \textbf{Model} & \textbf{Tasks} & \textbf{Lat. (FP32)} & \textbf{Lat. (INT8)} & \textbf{Pwr-Norm. Lat. (FP32)} & \textbf{Pwr-Norm. Lat. (INT8)} & \textbf{Avg Pwr} \\
    \hline
    \hline
    GhostNetv2 + 3 parallel heads (\textbf{Ours}) & SSC & 56.77 ms & 34.67 ms & 48.66 ms & 29.72 ms & 6.0 W \\
    \hline
    DeepLabV3 + MobileNetV3-L (\textbf{DLMN}) & S  & 39.52 ms & 15.70 ms & 32.75 ms & 13.01 ms & 5.8 W \\
    \hline
    HRNet18 (\textbf{HR}) & S & 118.07 ms & 47.72 ms & 106.26 ms & 42.95 ms & 6.3 W \\
    \hline
    UNet + MobileViT Small (\textbf{mViT}) & S & 117.79 ms & 82.94 ms & 109.38 ms & 77.02 ms & 6.5 W \\
    \hline
    MobileNetV3 Large (\textbf{MN}) & C & 18.61 ms & 9.19 ms & 15.59 ms & 7.61 ms & 5.8 W \\
    \hline
    ResNet50 (\textbf{RN50}) & C & 45.98 ms & 15.00 ms & 42.04 ms & 13.71 ms & 6.4 W \\
    \hline
    ViT B-16 (\textbf{ViT}) & C & 364.79 ms & 296.27 ms & 343.94 ms & 279.34 ms & 6.6 W \\
    \hline
    \end{tabular}
    \end{center}
\end{table*}
\end{center}

\begin{center}
\begin{table*}[h]
    \centering
    \caption{Energy-Quality tradeoff on low-power hardware.}
    \label{tab:tradeoff}
    \begin{center}
    \begin{tabular}{c c c c c c c}
    \hline
    \textbf{Method} & \textbf{Latency (FP32)} & \textbf{E$_\Delta$ (FP32)} & \textbf{Q$_\Delta$ (FP32)} & \textbf{Latency (INT8)} & \textbf{E$_\Delta$ (INT8)} & \textbf{Q$_\Delta$ (INT8)}  \\
    \hline
    \hline
    \textbf{Ours} & 56.77ms & 0\% &  0\% & 34.67ms &  0\% &  0\% \\
    \hline
    \textbf{DLMN + DLMN + MN} & 97.65ms & +66.26\% & +2.82\% & 40.58ms &  +13.16\% &  -7.94\% \\
    \hline
    \textbf{HR + mViT + MN} & 254.46ms & +374.80\% & +8.68\% & 139.85ms & +329.32\% &  -14.58\% \\
    \hline
    \textbf{HR + HR + ViT} & 600.92ms & +1043.50\% & +3.46\% &  357.72ms & +1129.06\% & +5.59\% \\
    \hline
    \textbf{mViT + mViT + ViT} & 600.36ms & +1056.29\% & +4.45\% & 462.15ms & +1358.31\% &  -21.73\% \\
    \hline
    \textbf{DLMN + DLMN + RN50} & 125.01ms & +120.95\% & +2.23\% & 46.39ms & +33.68\% &  -16.84\% \\
    \hline
    \end{tabular}
    \end{center}
\end{table*}
\end{center}

\begin{table*}
\centering
\caption{Backbone ablation}
\label{table:backbone}
\begin{tabular}{ccccc}
\hline
\textbf{Backbone}                    & \textbf{Cloud (mIOU)} & \textbf{Floods (mIOU)} & \textbf{Marine Litter (mF1)} & \textbf{Latency} \\
\hline
\hline
GhostNetV2 (SSL) + 3 parallel heads  & 82.33 & 40.32 & 68.03 & 56.77 ms \\
\hline
MobileNetV3-L (SSL) + 3 parallel heads & 81.45 & 40.00  & 64.78  & 32.51 ms  \\
\hline
\end{tabular}
\end{table*}

\subsection{Analysis on low-power hardware}

In order to validate the suitability of the proposed design for onboard usage, we performed some tests on an Nvidia Jetson Orin Nano 8GB embedded system. While not currently space-qualified, it is a low-power hardware platform with a CPU and GPU for AI acceleration with a peak power budget of 7W or 15W, depending on usage mode, which allows us to characterize latency, total energy consumption as well as limitations in image size due to memory on a sufficiently representative system. All tests are conducted in the 7W board mode. 

Table \ref{tab:speed_jetson} reports some results for a $512\times512\times4$ input in terms of inference latency, average power consumption and power-normalized latency. The latter is computed as the product between latency and average power normalized by 7W, i.e. the maximum power budget of the system. It should be noticed that average power consumption serves as a validation of whether the method is fully utilizing the available resources, by staying close to the 7W budget, or not. Latency results are averaged over 10 runs with 200 warmup iterations. We compare the proposed multitask architecture with the aforementioned baseline and state-of-the-art architectures for individual tasks. We can notice that the proposed design can solve three tasks (two segmentation (S) and one classification (C))  with a latency that is inferior of several other single-task models. We also notice that INT8 quantization provides almost a factor of 2 speedup. Considering the input resolution has $512\times512$ spatial locations to be labeled, we can say that the FP32 inference time of 56.77 ms corresponds to a throughput of 4.62 Mpx/s and that the INT8 inference time of 34.67 ms corresponds to a throughput of 7.56 Mpx/s.

Table \ref{tab:tradeoff} presents a tradeoff analysis for the multitask problem. In this analysis, we investigate what is the total energy consumed to solve the three tasks, as a function of latency and instantaneous power, in relation to the inference quality. The proposed design is compared with different combinations of methods to address the three tasks. These methods need to be run serially as the system is already fully used by each single task. In particular, we select some interesting combinations including fastest baseline (DLMN + DLMN + MN, refer to Table \ref{tab:speed_jetson} for acronyms), highest FP32 quality (HR + mViT + MN), highest INT8 quality (HR + HR + ViT), Transformers-only (mViT + mViT + ViT), and CNN-only (DLMN + DLMN + RN50). Taking the proposed design as the reference, we report $E_\Delta$ as the percentage difference between the energy in Joules consumed to complete the three tasks, and $Q_\Delta$ as the average percentage difference in inference metrics (i.e., the percentage difference in mIOU or F1 is computed for each task and then averaged over the three tasks). We are not surprised that because of the parallel multitask approach, the proposed design requires the least energy to complete the tasks. While it does not provide the best quality overall, modest improvements in quality are offset by large increases in energy consumption (e.g., +8\% quality requires +374\% energy with HR + mViT + MN), highlighting the good trade-off achieved by the proposed method.

\begin{figure}
\centering
\includegraphics[width=0.9\columnwidth]{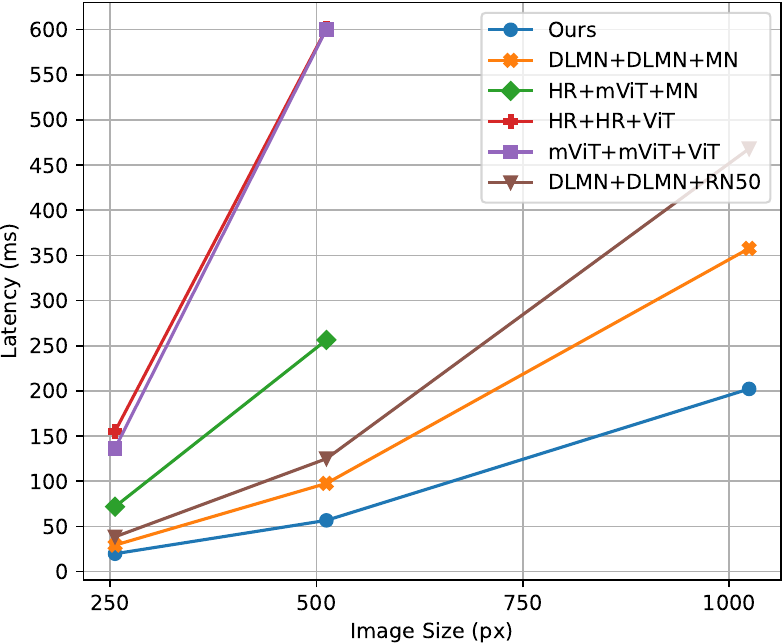}
\caption{Inference times using different image sizes on FP32 models. Pipelines that include mViT or ViT have no latency for 1024$\times$1024 images because there is insufficient RAM on the Jetson system to run them.}
\label{fig:4.1}
\end{figure}

\begin{figure}
\centering
\includegraphics[width=0.9\columnwidth]{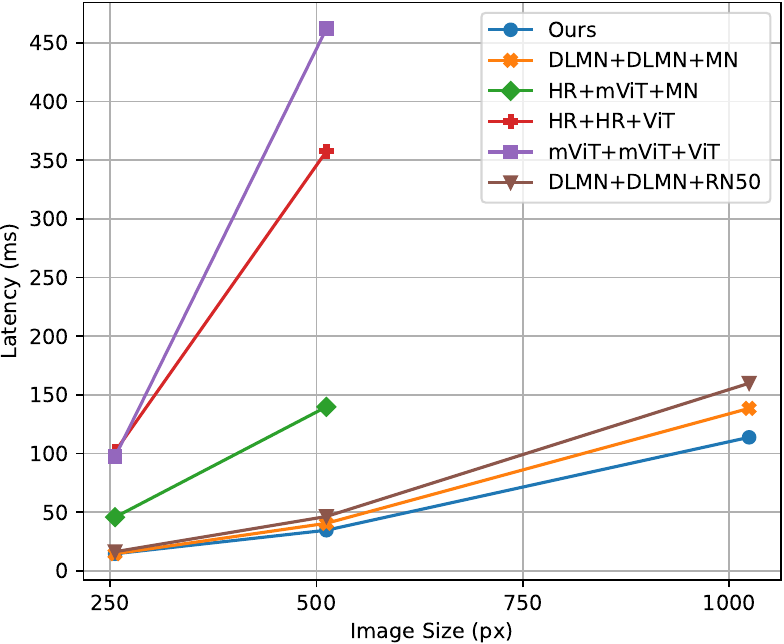}
\caption{Inference times using different image sizes on INT8 models. Pipelines that include mViT or ViT have no latency for 1024$\times$1024 images because there is insufficient RAM on the Jetson system to run them.}
\label{fig:4.2}
\end{figure}

Finally, we present how the methods scale as a function of image size in Fig. \ref{fig:4.1} and Fig. \ref{fig:4.2}. The upper limit in image size, dictated by the Jetson's 8GB shared system memory, is $1024\times1024$ for all methods except Transformers, which run out of memory at this resolution. Generally, slightly better efficiency is achieved with a $1024\times1024$ input, reaching 8.91 Mpx/s of INT8 throughput compared to 7.56 Mpx/s for a $512\times512$ input. Indeed, real onboard acquisitions may be significantly larger than $512\times512$ or $1024\times1024$. However, a tiling strategy would be adopted onboard, where the large image would be partitioned into tiles as large as the system memory allows. This is also why we present results in terms of throughput, which normalizes latency by image size.

\subsection{Backbone Architecture Ablation}

The choice of using GhostNetV2 as the backbone architecture for our main experiments was driven by being the state-of-the-art model among low-complexity architectures. However, one might wonder how a different backbone compares to GhostNetV2 under the specific SSL multitask setting under study. For this purpose, we test MobileNetV3-L as an alternative and present the results in Table \ref{table:backbone}. It can be noticed that MobileNet offers a different tradeoff between accuracy and latency, being worse on the former and faster for the latter. Whether this is desirable, it depends on the specifics of the mission under design, and, in particular, its speed target.

For a fair comparison, both GhostNetV2 and MobileNetV3-L were trained under identical conditions: the SSL pre-training was conducted for the same number of epochs on the same dataset, using a width multiplier of 1.6 for both networks. Furthermore, the application-specific heads employed for evaluation were consistent with those described in Subsection \ref{subsec:heads}.

\subsection{Backbone Size Ablations}

As mentioned above, all experiments conducted on a GhostNetV2 backbone with the $\alpha$ width parameter set to 1.6. Since this parameter influences the number of features in different layers of the network and the input channels for the various heads, we conducted experiments to explore how the network's parameters, MACs, FLOPs, and performance change with varying $\alpha$ values, reported in Table \ref{tab:modelsize}. In this experiment, SSL training for the backbone feature extractor for 100 epochs is followed by supervised training of the segmentation head for the cloud cover task. It is interesting to notice that the model scales beyond the $\alpha=1.6$ value used in all our experiments with improved mIOU. However, complexity also scales accordingly, so, while we found $\alpha=1.6$ to be a good tradeoff, if a mission desires it can sacrifice some speed for higher quality maintaining the proposed design by choosing $\alpha=2$ or higher.

\begin{table}[t]
    \centering
    \caption{Model size ablation for cloud cover segmentation.}
    \label{tab:modelsize}
    \setlength{\tabcolsep}{3pt}
    \begin{center}
    \begin{tabular}{c c c c c c}
    \hline
    \textbf{width $\alpha$} & \textbf{Params} & \textbf{MACs} & \textbf{FLOPs} &  \textbf{mIoU (FP32)} & \textbf{mIoU (INT8)} \\
    \hline
    \hline
    1 & 3.78M & 912.28K & 1.89G & 79.86 & 79.54\\
    \hline
    1.6 & 9.55M & 2.18G & 4.47G & 80.93 & 80.19\\
    \hline
    2 & 14.69M & 3.34G & 6.81G & 81.83 & 81.05\\
    \hline
    \end{tabular}
    \end{center}
\end{table}

\subsection{Comparison of Self-Supervised Learning Methods}

We conducted a comparative analysis of our local contrastive loss technique, detailed in Sec. \ref{subsec:backbone}, against the method presented in \cite{ssl_remote_sensing}, which was specifically tailored for contrastive learning training on remote sensing RGB images and can be considered a state-of-the-art self-supervised training approach in the remote sensing field. 

In order to provide a fair assessment of the training procedure, we used the same GhostNetV2 backbone architecture of our main experiments. The original authors code was adapted to handle transformations over 4 spectral channels, while we preserved the original hyperparameters and experimental settings as in the original work, pretraining the backbone on BigEarthNet dataset. Then, we froze the weights of our GhostNetV2 backbone and conducted a comprehensive evaluation on the cloud cover dataset, replicating the methodology outlined in \ref{subsec:performances}. The results are summarized in Table \ref{tab:ssl} and clearly demonstrate the substantial performance enhancement achieved through our self-supervised learning pretraining compared via a combination of local and global contrastive loss, with respect to the methodology proposed in \cite{ssl_remote_sensing}.

\begin{table}[t]
    \centering
    \caption{SSL Training Ablation for cloud cover segmentation.}
    \label{tab:ssl}
    \setlength{\tabcolsep}{3pt}
    \begin{center}
    \begin{tabular}{c c}
    \hline
    \textbf{SSL Training} & \textbf{mIoU (FP32)} \\
    \hline
    \hline
    Local Contrastive Loss (ours) & 82.33  \\
    \hline
    SSL Remote Sensing \cite{ssl_remote_sensing} & 79.16 \\
    \hline
    \end{tabular}
    \end{center}
\end{table}

\section{Conclusions}

We presented a high-level conceptualization of how to design an AI payload for a spacecraft capable of addressing multiple tasks of interest directly onboard to provide rapid response to events or improved system functionality. We also delved into a low-complexity architecture and its training process, leveraging self-supervised learning to enable a modular approach as well as reduce requirements for labeled data. Extensive experiments over three tasks of interest on low-power hardware show that the proposed method is capable of inference quality close to that of high-complexity state-of-the-art models at a fraction of energy consumption. Moreover, we measured a high absolute throughput that would make real-time operations feasible.

%\clearpage
\bibliographystyle{IEEEtran}
\bibliography{biblio}

\end{document}